# Monolayer Graphene as Saturable Absorber in Mode-locked Laser

Qiaoliang Bao[†,∥], Han Zhang[‡,∥], Zhenhua Ni[§], Yu Wang[†], Lakshminarayana Polavarapu[†], Zexiang Shen[§], Qing-Hua Xu[†], Ding Yuan Tang[‡,*], and Kian Ping Loh[†,*]

[†] Department of Chemistry, National University of Singapore, 3 Science Drive 3, Singapore 117543

[‡] School of Electrical and Electronic Engineering, Nanyang Technological University, Singapore 639798

[§] School of Physical and Mathematical Sciences, Nanyang Technological University, Singapore 637371

Page Numbers. The font is ArialMT 16 (automatically inserted by the publisher)

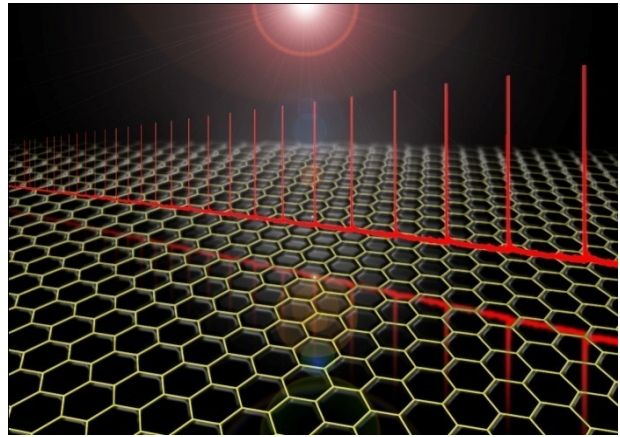

The absorption of one atomic layer graphene can be saturated at quite low excitation intensity and it gives a remarkably large modulation depth of 65.9%. Picoseconds laser pulses (1.23 ps) can be generated using monolayer graphene as saturable absorber. Compared to multilayer graphene, monolayer graphene mode-locked laser shows better pulse stability and output energy.

# Monolayer Graphene as Saturable Absorber in Mode-locked Laser

Qiaoliang Bao[1, †], Han Zhang[2, †], Zhenhua Ni[3], Yu Wang[1], Lakshminarayana Polavarapu[1], Zexiang Shen[3], Qing-Hua Xu[1], Ding Yuan Tang[2](✉), and Kian Ping Loh[1] (✉)

[1] Department of Chemistry, National University of Singapore, 3 Science Drive 3, Singapore 117543
[2] School of Electrical and Electronic Engineering, Nanyang Technological University, Singapore 639798
[3] School of Physical and Mathematical Sciences, Nanyang Technological University, Singapore 637371
[†]: These authors contributed equally to this work.

## ABSTRACT

We demonstrate that the intrinsic properties of monolayer graphene allow it to act as a more effective saturable absorber for mode-locking fiber lasers compared to multilayer graphene. The absorption of monolayer graphene can be saturated at lower excitation intensity compared to multilayer graphene, graphene with wrinkle-like defects, and functionalized graphene. Monolayer graphene has a remarkable large modulation depth of 65.9%, whereas the modulation depth of multilayer graphene is greatly reduced due to nonsaturable absorption and scattering loss. Picoseconds ultrafast laser pulse (1.23 ps) can be generated using monolayer graphene as saturable absorber. Due to the ultrafast relaxation time, larger modulation depth and lower scattering loss of monolayer graphene, it performs better than multilayer graphene in terms of pulse shaping ability, pulse stability and output energy.

## 1. Introduction

Graphene is an atomic layer of conjugated $sp^2$ carbon atoms arranged in a two dimensional hexagonal lattice, charge carriers in it move at ultrafast speed behaving like relativistic, massless Dirac particles [1,2]. Most of the research attention thus far has been focused on the unique electronic structure of graphene. Comparatively less effort has been dedicated to investigate the photonic properties and applications of graphene, although graphene shows exciting potential in ultrafast photonics devices because of the ultrafast carrier dynamics [3,4] and large, broadband optical

Address correspondence to K. P. Loh, chmlohkp@nus.edu.sg; D. Y. Tang, edytang@ntu.edu.sg.

absorption (2.3 % per layer) [5,6]. Previously, we have demonstrated the applications of few layers graphene [7-9] as well as graphene-polymer composite [10,11] as saturable absorbers in mode-locked lasers. Following, another two groups also confirmed the mode-locking of lasers using chemically processed graphene film [12] and graphene-polymer composites [13] as saturable absorbers. In all these works, the saturable absorbers are multilayer graphene films or graphene composites. The unambiguous demonstration of saturable absorption from single atomic layer of graphene film has yet to be achieved. In principle, the excited carriers in pristine monolayer graphene are expected to show ultrafast decay (similar to graphite in tens femto-second [14]) due to the zero gap in graphene. It therefore raises a fundamental question: can the excited carriers in pristine monolayer graphene fill all the states in conduction band or reach equilibrium under such a fast carrier relaxation rate? It must also be recognized that the optical and electrical properties of atomically thin graphene will be very sensitive to defects. The influence of sample quality and surface states on the saturation absorption properties has not been investigated.

In this work, in contrast to all previous studies that examine composite polymer or multilayer films involving graphene, we have painstakingly isolated a single atomic layer of graphene on an optical fiber and studied its nonlinear absorption properties. The performance is compared with monolayer graphene with surface defects, or monolayer graphene modified with functional groups, as well as multilayer graphene. We found that the threshold energy for saturation correlates with nonsaturable loss and scattering effects, all of which scale with defects or thickness of the film. In addition, monolayer graphene is found to have a faster carrier-carrier and carrier-phonon intraband scattering process than multilayer graphene.

## 2. Experimental

Large area monolayer and multilayer graphene films were grown by CVD on Cu [15] and Ni respectively. Cu and Ni [16] were then etched away in FeCl3 solution, followed by the transfer of the graphene films onto optical fibers or quartz substrate, as described in our previous work [7]. Monolayer graphene was functionalized covalently with nitrobenzene by immersing the end face of fiber pigtail into 4-Nitrobenzenediazonium tetrafluoroborate (Sigma-Aldrich) solution (0.01 mg •mL$^{-1}$) for 10 min. The sample was then washed by deionized water. The Raman spectra and images were measured on WITEC CRM 200 Raman system (532 nm, 100 × objective lens). The time-resolved pump-probe profiles of graphene films were obtained using a femtosecond Ti:sapphire laser system (Spectra Physics). The laser pulses were generated from a mode-locked Ti:sapphire oscillator seeded regenerative amplifier with a pulse energy of 2 mJ, pulse width of 100 fs at 800 nm and a repetition rate of 1 kHz. The 800 nm laser beam was split into two portions. The larger portion of the beam was directed through a BBO crystal to generate the 400 nm pump beam by frequency-doubling. A small portion of the 800 nm pulses was used to generate a white light continuum probe beam in a 1 mm sapphire plate. The signal and reference beams were detected by photo-diodes that were connected to lock-in amplifiers and the computer. The pump beam was focused onto the film with a beam size of 200 $\mu$m and overlapped the smaller-diameter (100 $\mu$m) probe beam. The delay between the pump and probe pulses was varied by a computer-controlled translation stage. The pump beam was modulated by an optical chopper at a frequency of 500 Hz.

For the power-dependent nonlinear absorption measurements, a stable and standard soliton mode locked fiber laser working at 1550 nm with output pulse width of ~1 ps and repetition rate of ~5 MHz was used as input seed pulse, similar to our previous work [7,11]. The measurements started at a very low input power of ~ -40 dBm (i.e., 100 nW). After the seed pulses pass through monolayer graphene, an output power of ~ -45 dBm (31.6 nW) was observed. We then gradually increased the input power to -14 dBm or so, the output power gave a value close to that with about 0.15 dB transmission loss due to saturation. For application of graphene-based fiber lasers, we used the standard fiber-optic components such as WDM, PC, coupler, optical isolator, EDF and SMF. The fiber laser has a ring cavity which consists of a piece of 6.4 m EDF with a group velocity dispersion (GVD) parameter of 10 (ps•nm$^{-1}$)•km$^{-1}$, and a total length of 72.2 m SMF with GVD parameter of 18 (ps•nm$^{-1}$)•km$^{-1}$. Polarization independent isolators were used to force the unidirectional operation of the ring, and an intra-cavity PC was used to optimize the cavity birefringence. The laser was pumped by a high power Fiber Raman Laser of wavelength 1480 nm.

## 3. Results and discussion

### 3.1. Characterizations of monolayer graphene

The monolayer graphene was synthesized by the CVD approach on Cu substrate [15], in which uniform, high quality mono-layer graphene was obtained through the catalytic decomposition of CH$_4$ on the Cu surface [17]. Multilayer graphene (1~4 layers) grown on Ni was also studied for comparison [7]. The graphene film (5 mm × 5 mm) was isolated by etching off the Cu (or Ni) substrate and then it was transferred onto the cross-section of optical fiber for further studies.

To understand the influences of defects, functional groups and number of layers on the nonlinear optical properties, the following samples were specially prepared for investigation: monolayer graphene, monolayer graphene covalently bonded to functional groups on its basal plane by diazonium functionalization, monolayer graphene with wrinkles, as well as 1~4 layers and 4~8 layers graphene grown on Ni. Raman spectroscopy was used to evaluate the quality and uniformity of graphene in fiber core area, as shown in Figure 1. The location of the ~ 8 μm fiber core is determined by Raman signal in the 200 - 600 cm$^{-1}$ range, which originates from crystalline $SiO_2$ core, described elsewhere [7]. Figure 1a shows that monolayer graphene film covers quite homogenously on the fiber core-area. The corresponding Raman spectrum in Figure 1e reveals that the Raman 2D band is much stronger than G band with a 2D/G ratio of 2.4 and the 2D band can be fitted well by a single Lorentzian curve with a narrow full-width at half-maximum (FWHM) of 28 cm$^{-1}$. The Raman D band associated with defect is hardly observed [18]. These indicate that the as-produced graphene has very high crystalline quality. The atomic force microscopy (AFM, Dimension 3100 SPM) image further confirms the monolayer thickness of the produced graphene, ~ 0.7 nm (see Figure S1 in Electronic Supplementary Material). Figure 1b reveals that the functionalized monolayer graphene is generally uniform except for the presence of several wrinkles on the edge area of fiber core. The corresponding Raman spectrum in Figure 1e exhibits characteristic peaks of both graphene (G and 2D bands) and 4-Nitrophenyl groups in 1000~1500 cm$^{-1}$ region, which suggests successful functionalization *via* covalent bonds. Figure 1c shows that the graphene surface comprises many wrinkle-like defects, these manifest as strong D band [19] in the corresponding Raman spectrum in Figure 1e. The multilayer graphene (grown on Ni) is not uniform in thickness, as shown in Figure 1d. The relative broad 2D band in its Raman spectrum can be fitted by several Gaussian or Lorentzian peaks, which is characteristic of its few layered thickness.

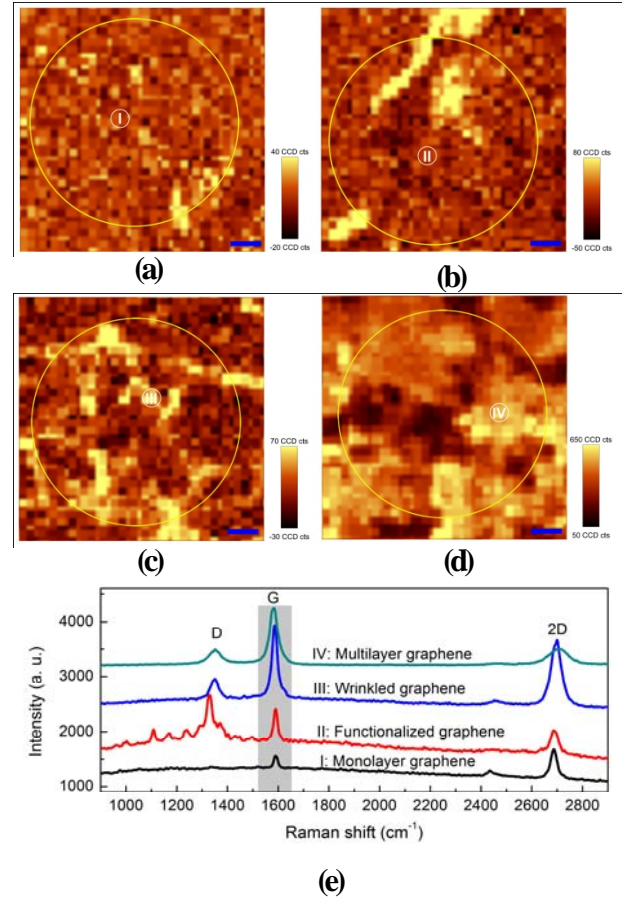

**Figure 1** Micro-Raman characterization of monolayer and multilayer graphene transferred onto optical fiber. Integrated intensity maps of G band (1520-1650 cm$^{-1}$) of (a) monolayer graphene, (b) functionalized monolayer graphene, (c) wrinkled monolayer graphene and (d) multilayer graphene. The yellow circles indicate the location of fiber core area of each sample. All the scale bars: 1 μm. (e) Raman spectra from marked spots with corresponding number in (a-d).

### 3.2. Nonlinear optical properties

The saturable absorption of the graphene samples was investigated and presented in Figure 2. Figure 2a shows the nonlinear absorption properties (in units of dB) of monolayer graphene with different surface states. First, it can be seen that the absorption of all the samples decreases sharply at a particular threshold input intensity and becomes nearly constant at higher intensity. This reflects typical saturable absorption feature. It is noteworthy that pristine monolayer graphene (the sample in Figure 1a) only causes 0.14 dB transmission loss after its absorption was saturated, in contrast to more than 5 dB initial optical loss upon 1 ps laser pulses at lower incident intensity. Notwithstanding the insertion loss of the fiber pigtail (~2.3 dB), this result reveals that graphene becomes highly transparent after it is saturated. The presence of wrinkles (the sample in Figure 1c) in graphene results in slightly higher transmission loss after saturation (inset of Figure 2a), this may originate from nonsaturable optical loss. It is also found that the transmission loss in saturation regime decreases a little bit after the sample was covalently functionalized (the sample in Figure 1b), which might correlate with p-type doping by 4-Nitrophenyl groups.

For the purpose of comparison, the nonlinear saturable absorption of few-layer graphene was also measured, as shown in Figure 2b. Thicker graphene films have larger optical loss in both linear and nonlinear regime. For 1-4 layers graphene (the sample in Figure 1d), the transmission loss decreases from 5.88 to 1.65 dB with increasing incident intensity. It decreases from 8.55 to 4.92 dB for the 4-8 layers graphene sample. The performance of pristine monolayer graphene contrasts sharply with these because the transmission loss decreases dramatically from 5.10 to 0.14 dB after saturation. These results point to higher nonsaturable loss caused by the non-uniformity and scattering of thicker graphene films. We should also note that the whole nonlinear absorption curves shift in the direction of higher input intensity for thicker films, which reflects that higher input power is needed to saturate all the graphene layers in the longer optical absorption path. The strong scattering has also contributed to this shift as some of the input power is wasted.

The nonlinear absorption data collected above was converted into percent transmission so that we can clearly compare the two components of nonlinear absorption: saturable absorption and nonsaturable absorption, as shown in Figure 2c. The saturable absorption is also termed as modulation depth, which is very important to evaluate the pulse shaping abilities of saturable absorbers. The saturation intensity, i.e., optical intensity required in a steady state to reduce the absorption to half of its unbleached value, can be estimated by fitting the nonlinear absorption curves using formula

$$\alpha(I) = \frac{\alpha_s}{1 + I/I_S} + \alpha_{NS} \qquad (1)$$

where $\alpha_S$ and $\alpha_{NS}$ are the saturable and nonsaturable absorption, $I_s$ is the saturation intensity [11]. These saturable absorption properties are summarized in Table 1.

The remarkable result is that pristine monolayer graphene has a modulation depth of 65.9%, which is much larger than those reported in previous mode locking studies on graphene [7,8,10,11] as well as those of any other known saturable absorbers [20,21]. It can be concluded that both surface defects (wrinkles) and thickness will result in smaller modulation depth due to the increased nonsaturable absorption. After functionalization, the modulation depth is enhanced a little bit from 59.7 % to 64.4 %. The saturation intensity shows the opposite trend versus surface defects and thickness. The pristine monolayer graphene has the smallest saturation intensity of 0.53 MW•cm$^{-2}$, while two small wrinkles on the surface (shown in Figure 1b) leads to 4% increase of the saturation intensity. The 1-4 layers graphene film has a saturation intensity of 0.77 MW•cm$^{-2}$, which agrees well with previous work [7]. Scaling with thickness, the 4-8 layers graphene needs much more incident light power to reach saturation with a saturation intensity of 1.09 MW•cm$^{-2}$. We suggest that the defect induced nonsaturable loss and scattering play a dominant role to affect the saturation intensity.

### 3.3. Carrier dynamics

To understand the saturation mechanism, especially

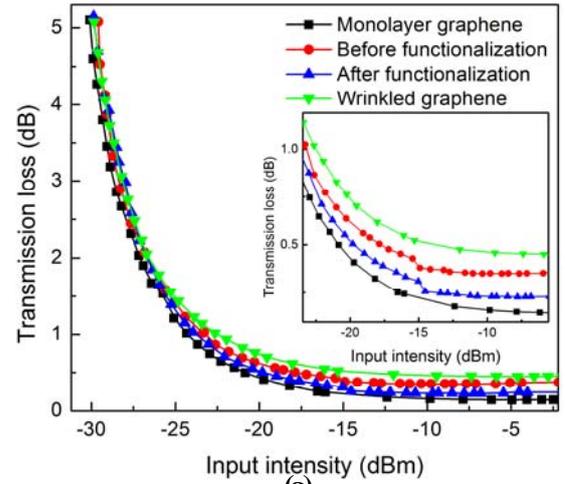

(a)

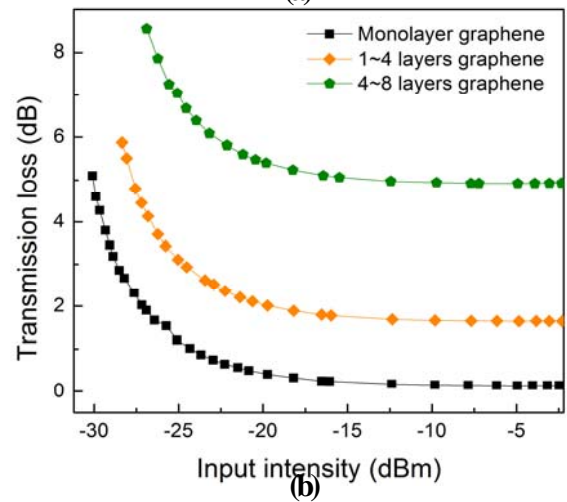

(b)

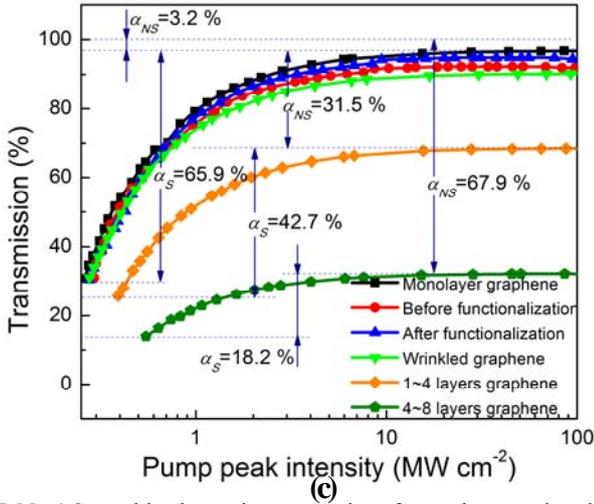

**Figure 2** Power-dependent nonlinear absorption properties. (a) Monolayer graphene samples with different surface states. Inset shows enlarged plots. (b) Comparison of monolayer graphene with multilayer graphene. (c) Power-dependent percent transmission.

**Table 1** Saturable absorption properties of monolayer and multilayer graphene

| Sample number | Transmission loss | | Saturation intensity | Normalized absorbance | |
|---|---|---|---|---|---|
| | Starting point | After saturation | | $α_S$ | $α_{NS}$ |
| | (dB) | (dB) | (MW·cm$^{-2}$) | (%) | (%) |
| 1 | 5.10 | 0.14 | 0.53 | 65.88 | 3.24 |
| 2 | 5.08 | 0.35 | 0.55 | 59.65 | 9.30 |
| 3 | 5.15 | 0.23 | 0.57 | 64.36 | 5.09 |
| 4 | 5.08 | 0.45 | 0.54 | 59.02 | 9.90 |
| 5 | 5.88 | 1.65 | 0.77 | 42.69 | 31.48 |
| 6 | 8.55 | 4.92 | 1.09 | 18.17 | 67.85 |

1: Monolayer graphene. 2: Monolayer graphene with two wrinkles before functionalization. 3: Monolayer graphene with two wrinkles after functionalization. 4: Monolayer graphene with many wrinkles. 5: 1~4 layers graphene grown on Ni. 6: 4~8 layers graphene grown on Ni with respect to the dynamics of state filling by the photocarriers which lead to Pauli blocking, pump-probe experiments were carried out. Figure 3a, b show the measured transmittivity transients of pristine monolayer graphene and 1-4 layers graphene, which is pumped at 400 nm and probed at 750 nm. The data is well fitted using a bi-exponentially decaying function, $\Delta T(t)/T = A_1 \exp(-t/\tau_1) + A_2 \exp(-t/\tau_2)$, convoluted with the cross-correlation of the pump and probe pulses. The carrier relaxation time comprises a fast time constant ($\tau_1$) in the 100 - 150 fs range, resulting from carrier-carrier intraband scattering process, and a slower time constant ($\tau_2$) in the 405 - 570 fs range, corresponding to carrier-phonon intraband scattering and electron-hole recombination (Auger scattering). The initial fast relaxation time $\tau_1$ of monolayer graphene is around 100 fs, which agrees with previous reports on graphene [3,22-24]. However, we should note that the faster time $\tau_1$ is of the order of the pulse width and is, therefore, not accurately resolved. Compared with the theoretical prediction by Rana et al. [25], the electron-hole recombination rate ($\tau_2$) is quite fast (< 1 ps) in our experimental observation. This can be explained by the dependence of charge screening and Coulomb scattering on the dielectric constant of the surrounding media. Contrary to Dawlaty's
7

observation that larger crystal disorder results in shorter relaxation time ($\tau_2$) in expitaxial graphene [22], we observed that the 1-4 layers graphene has longer relaxation time (both $\tau_1$ and $\tau_2$) than monolayer graphene even though it has stronger defect-related D peak (it might also originate from the edges of inhomogeneously stacked graphene layers) in its Raman spectrum. This trend is however consistent with Newson's observations of the layer dependence of carrier relaxation decay time. [23]

We can estimate the saturated carrier density using simplified rate equation which involves long-pulse excitation [7]

$$N = \frac{\alpha I \tau}{\hbar \omega} \qquad (2)$$

where $\alpha$ is the absorption coefficient, $I$ is the incident intensity, $\tau$ is the carrier recombination time, $\hbar$ is reduced Planck's constant and $\omega$ is the frequency of light. $\hbar\omega$ = 0.8 eV as the pump laser for saturable absorption measurements is at 1550 nm. For monolayer graphene, let's assume that $\tau \approx \tau_2$ = 405 fs, $I_s$ = 0.53 MW·cm$^{-2}$, it gives saturated carrier density $N_s$ = 1.06 × 10$^{13}$ cm$^{-2}$. According to the relationship between Fermi energy and carrier density at low temperature [26,27],

$$E_F(n) = \hbar|v_F|\sqrt{\pi n} \qquad (3)$$

where $|v_F|$ is the Fermi velocity and $n$ is the carrier density in units of cm$^{-2}$, we roughly estimate that these amount of carriers can fill the cone-shaped conduction band of graphene up to 0.43 eV (this value is over-estimated as the spread of carrier distribution at finite temperature is ignored), which is very close to the electron energy excited by our pump laser, which is $\hbar\omega/2$ = 0.4 eV. Therefore, we can conclude that all the electron states below $\hbar\omega/2$ have been occupied by pumped electrons, prohibiting the optical transitions and further creation of electon-hole pairs due to Pauli blocking, as schematically illustrated by inset of Figure 3c.

Energy dependent relaxation time measurements were carried out by varying the probe laser wavelength, as shown in Figure 3c, in which the trace of graphite extracted from reference [14] was also included for comparison. It is found that the measured lifetime of graphene is longer than graphite as the faster relaxation time $\tau_1$ is not resolvable, limited by the pulse width used in our measurements. The quasi-particle lifetime in graphite was theoretically and experimentally demonstrated to be inversely proportional to the quasiparticle energy above the Fermi level[14,28], due to the linear dispersion around Fermi level, even though some reports discovered deviations between 1.1 and 1.5 eV [29,30]. The key point of Figure 3c is that it suggests that monolayer and multilayer graphene follow the general trend in graphite, i.e., lower excitation energy correlates to longer carrier lifetime. At the working wavelength of telecommunication C band (~1550 nm), the lifetime should be longer than 100 fs in monolayer graphene. Similar to conventional semiconductor absorbers, the slower relaxation time constant ($\tau_2$) results in reduced saturation intensity for part of

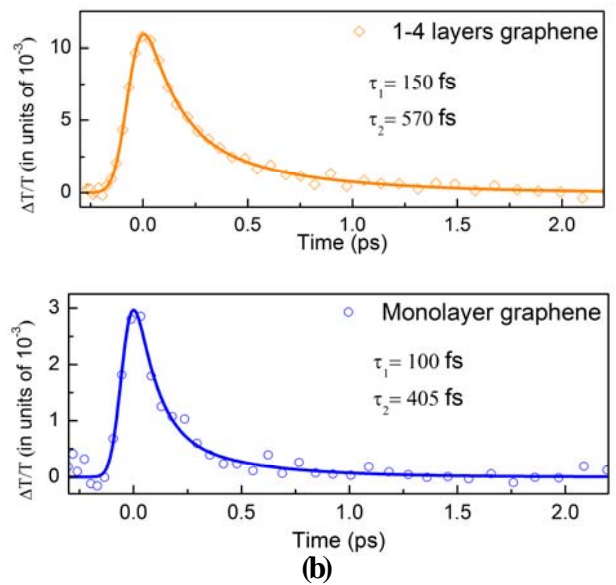

**(b)**

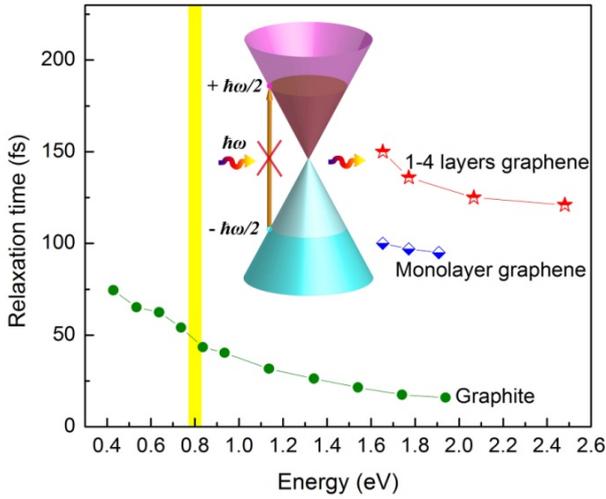

**Figure 3** Measured transmittivity transients of graphene samples. (a) 1-4 layers graphene. (b) Monolayer graphene. The samples were pumped at 400 nm and probe at 750 nm with 20mW incident laser power. (c) Energy dependant relaxation time ($\tau_1$) measurements pumped at 400 nm and probe at 500, 600, 700 and 750 nm. The green solid dot curve was adapted from reference [14] from comparison. The yellow region indicates the operating window (~ 1550 nm, 0.8 eV) for telecommunication C band. Inset shows a schematic of saturable absorption in graphene due to Pauli blocking.

the absorption and hence facilitates self-starting mode-locking. In contrast, faster relaxation time ($\tau_1$) constant is more effective in shaping ultra-short laser pulses [20,21]. Accordingly, monolayer graphene should have better pulse shaping performance than multilayer graphene due to shorter relaxation time ($\tau_1$) at the scale of ~100 fs.

### 3.4. Pulse shaping ability

To evaluate the pulse shaping ability of graphene and correlate this with its carrier relaxation time ($\tau_1$), we carried out pulse dynamics simulation. As mentioned, the absorption in graphene can be saturated under strong excitation due to the depletion of final states (i.e., Pauli blocking). Within 100 fs, the excited carriers in each band thermalize, which leads to partial recovery of the absorption. Such carrier dynamics in graphene influences the intra-cavity mode-locked pulse behaviors, i.e., longer pulses transmitted through graphene would encounter further pulse shaping (the lower intensity wing will be absorbed), owing to graphene's fast relaxation time. In order to quantitatively characterize the pulse shaping ability of graphene, comprehensive numerical simulations on the pulse dynamics in the laser cavity was carried out here based on the coupled Ginzburg-Landau equations (see Electronic Supplementary Material and Ref. [31,32]). The advantage of this simulation is that one can get an insight on the pulse evolution in the cavity, especially the pulse shaping caused by each of the cavity components, which are not easily visualized in experiments.

Under stable mode locking operation, the pulse durations before and after the graphene were compared, as shown in Figure 4. Accordingly, the ratio of pulse width variations was adopted to feature graphene's pulse shaping ability (See inset of Figure 4b). Figure 4a shows the calculated soliton pulse before and after graphene, obtained with relaxation time = 100 fs, gain = 15.5 dB and a very weak cavity birefringence of $L/L_b = 0.1$. It shows that before passing through graphene, the pulse width is about 1.93 ps. After passing through graphene it became about 1.66 ps, thus an obvious pulse width narrowing was observed. Considering that the

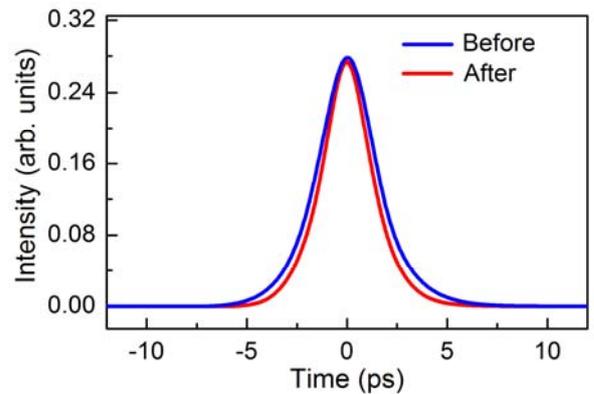

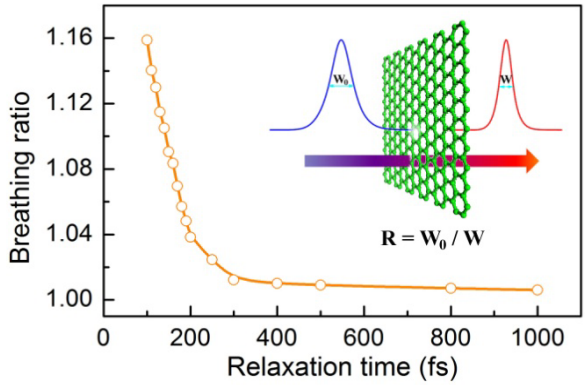

**Figure 4** (a) Calculated soliton pulse duration before and after it passes through graphene. (b) Breathing ratio of graphene as a function of relaxation time. Inset indicates the pulse shaping by graphene and definition of breathing ratio R = $W_0$ / W, where $W_0$ is the pulse width before graphene and W is the pulse width after graphene.

relaxation time of graphene could vary a lot from sample to sample due to defects, chemical functionalization and non-uniform thickness, we calculate the pulse width narrowing obtained using the same laser parameters but with different saturable relaxation times. Fig. 4b shows the result of simulated breathing ratio as a function of relaxation time. It shows that in a mode-locked fiber laser, the pulse shaping ability of graphene becomes less significant once its relaxation time becomes slower. This indicates that graphene with faster relaxation time possesses stronger pulse shaping ability, which points to the potential optoelectronic application of monolayer graphene as pulse shapers.

### 3.5. Mode-locked laser

Monolayer graphene was finally incorporated into a ring-configuration fiber laser cavity, as shown in Figure 5a. Figure 5b-f shows representative mode-locking characteristics of monolayer graphene. Figure 5b shows the oscilloscope trace of the laser output where a pulse repeats every ~393.3 ns, matching exactly with the cavity length ~78.6 m. In a laser cavity, the boundary condition causes light to be emitted at discrete frequencies, known as modes. When a saturable absorber material such as graphene is used to "mode-lock" these modes, the phases of all the modes become synchronized, and a stable and intense single bright pulse is formed (Figure 5b). The corresponding mode locking spectrum in Figure 5c has central wavelength located at 1561 nm and a 3 dB bandwidth of more than 5 nm. Experimentally, once the pump power is raised above threshold (~ 8 mW), we can always observe the mode locking spectra with clear and symmetric Kelly sidebands as shown in Figure 5c, which are caused by constructive interference between the soliton and dispersive waves emitted from the soliton. In such a soliton mode-locking regime, the function of saturable absorber (i.e., graphene) is to initiate pulse shaping and stabilize the pulse. Following, other laser cavity properties such as dispersion will play a more important role to determine the pulse parameters [33]. The formed soliton displays the nonlinear Schrödinger equation soliton features [32], in this case the total cavity dispersion is anomalous where intrinsic soliton shaping mechanism arises from the interplay between the anomalous cavity dispersion and fiber nonlinear optical Kerr effect.

The radio-frequency (RF) spectrum of the mode-locked pulses was also measured, as shown in Figure 5d, with a signal-to-noise ratio of 60 dB ($10^6$ contrast). It indicates that the laser is currently operated in an excellent CW mode-locking regime. An output power up to 3 mW was recorded with a slope efficiency of 2.2% (Figure 5e). The result is comparable to the SWNTs mode locked fiber lasers [34,35] but it could be further improved by laser cavity design, i.e., larger cavity output ratio and weaker splicing loss. The pump threshold for self-starting of the mode locking ranges from 8 to 40 mW, similar to multilayer graphene [7]. Since monolayer graphene saturable absorber has relatively lower insertion loss than multi-layer graphene, in principle, lower threshold pump energy is required to initial the mode locking performance. However, modulation depth and recovery time also plays an important role in determining the mode locking ability of saturable absorbers. Due to the fact that monolayer graphene has extremely large modulation depth together with fast decay time, higher pump is desirable to stabilize the mode locking operation. Consequently, we do not observe a significant change on the self-starting threshold of the monolayer graphene mode locking. The autocorrelation (AC) trace of the mode-locked pulses is shown in Figure 5f, which is well fitted by a sech$^2$ profile with 1.90 ps full-width at half-maximum (FWHM). The real pulse width is then obtained by multiplying the AC trace width with the de-correlation factor, which is 0.648 for the sech$^2$ pulses. This gives a pulse width of 1.23 ps, which is comparable with those of the pulses obtained in multilayer graphene [7] or SWNTs mode locked fiber lasers [36].

In a series of comparative mode locking studies using carbon nanotube saturable absorbers, we observed that pulse splitting is more likely to occur in the same laser cavity. In terms of laser performance, the larger modulation depth of monolayer graphene ensures long term stability of the generated laser pulses by avoiding pulse splitting. On the other hand, the small scattering loss favors the generation of high energy pulse in the same laser cavity design [37]. According to our preliminary experimental results, high energy mode locking up to 20 nJ was obtained by using monolayer graphene as saturable absorber in a normal dispersion laser cavity, which is much higher than that from multilayer graphene (7.3 nJ) in a similar cavity [8]. Furthermore, the mode locking performance does not degrade for a few weeks. This aspect will be further investigated.

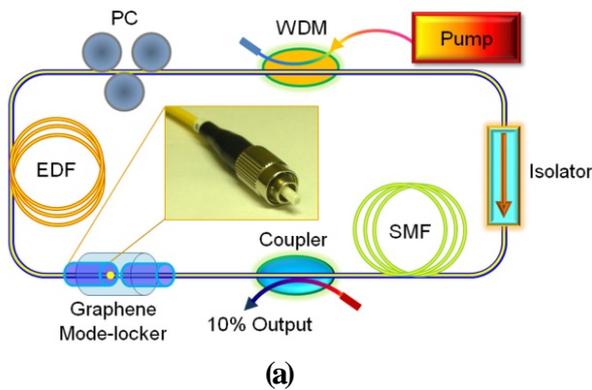

(a)

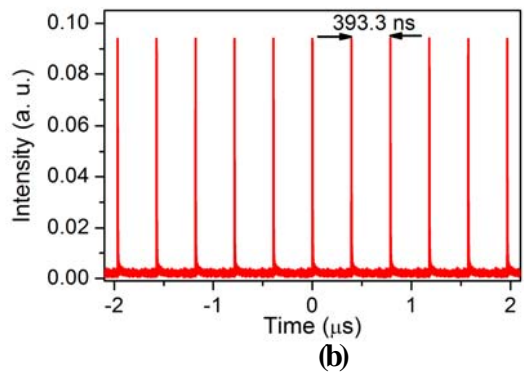

(b)

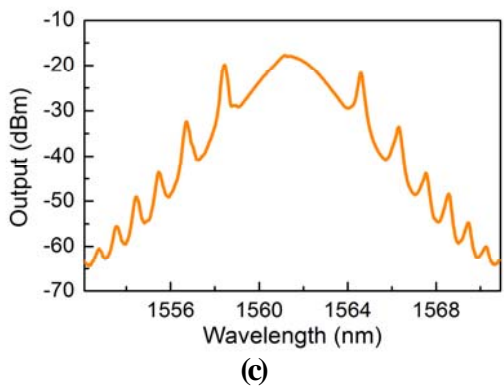

(c)

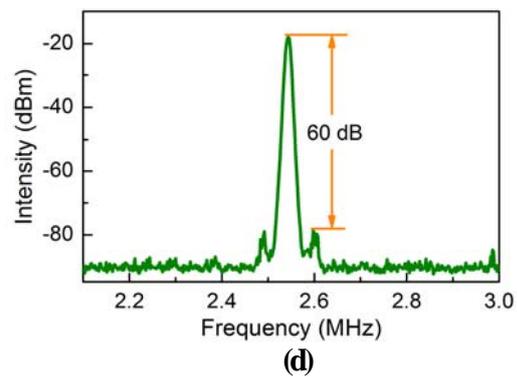

(d)

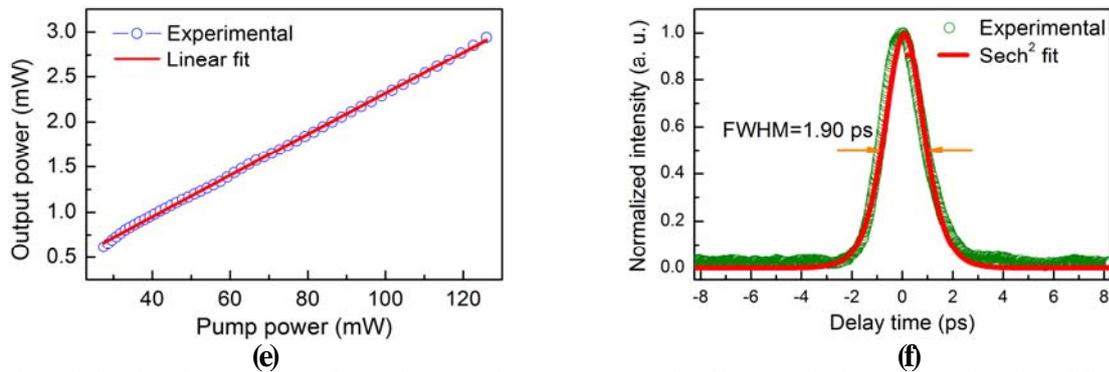

(e)                                (f)

**Figure 5** Mode-locking characteristics of monolayer graphene. (a) Schematic of laser cavity (WDM: wavelength division multiplexer; PC: polarization controller, EDF: erbium doped fiber; SMF: single mode fiber). Inset: photograph of fiber pigtail with graphene film coating on the end face. (b) Output pulse train. (c) Typical output spectrum centered at 1561 nm. (d) Radio frequency spectrum. (e) Output power as a function on pump power. (f) AC trace and $Sech^2$ fitting curve, 1.23 ps pulse is obtained.

## 4. Conclusions

We have comparatively evaluated the saturable absorption properties of monolayer graphene, multilayer graphene as well as defective graphene to gain an insight into how layer thickness and defects affect mode locking performance. In terms of saturation threshold energy, modulation depth and pulse shaping ability, monolayer graphene shows superior performance compared to thicker or defective graphene layers. Pulse dynamic simulation reveals that the pulse shaping ability is correlated with the ultrafast relaxation time of monolayer graphene. Monolayer graphene can be saturated at remarkably low excitation intensity of 0.53 MW•$cm^{-2}$ and its modulation depth of 65.9% is possibly the largest of all known saturable absorbers so far. We further demonstrated that picoseconds ultrafast laser pulse (1.23 ps) can be generated using monolayer graphene as saturable absorber.

## Acknowledgements

Q. Bao and H. Zhang contributed equally to the project. The authors thank the NRF-CRP grant "Graphene Related Materials and Devices" R-143-000-360-281.

**Electronic Supplementary Material**: Supplementary material (AFM image of monolayer graphene and pulse dynamics simulation) is available in the online version of this article at http://dx.doi.org/10.1007/10.1007/s12274-***-****-* (automatically inserted by the publisher) and is accessible free of charge.

# Electronic Supplementary Material

# Monolayer Graphene as Saturable Absorber in Mode-locked Laser

Qiaoliang Bao[1, †], Han Zhang[2, †], Zhenhua Ni[3], Yu Wang[1], Lakshminarayana Polavarapu[1], Zexiang Shen[3], Qing-Hua Xu[1], Ding Yuan Tang[2](✉), and Kian Ping Loh[1] (✉)

[1] Department of Chemistry, National University of Singapore, 3 Science Drive 3, Singapore 117543
[2] School of Electrical and Electronic Engineering, Nanyang Technological University, Singapore 639798
[3] School of Physical and Mathematical Sciences, Nanyang Technological University, Singapore 637371
[†]: These authors contributed equally to this work.

Address correspondence to K. P. Loh, chmlohkp@nus.edu.sg; D. Y. Tang, edytang@ntu.edu.sg.

## Atomic Force Microscopy

The atomic force microscopy (AFM) was carried out on Dimension 3100 SPM. The monolayer graphene sample was transferred onto SiO$_2$ substrate for AFM studies. Figure S1 shows typical tapping-mode AFM topography and cross-sectional profile. The transferring process causes folded edge and some wrinkles in graphene film. The measured thickness in flat area is ~ 0.7 nm, which confirms graphene's monolayer nature.

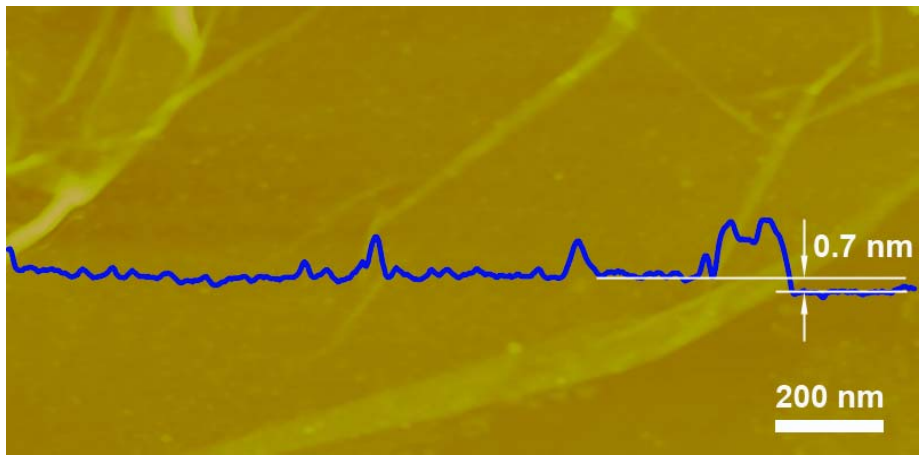

**Figure S1.** AFM topography and cross-sectional profile of monolayer graphene film. Vertical scale: 80 nm.

## Pulse Dynamics Simulation

In the present fiber laser mode-locked by saturable absorbers with relaxation time varying from 100 fs to 1 ps, soliton has been formed as judged from the presence of the spectral sideband on the optical spectrum and the small frequency chirp of the pulses. To correlate the modulation depth of different layers graphene with their mode locking behavior, the operation of the laser was numerically simulated.[1, 2] The following coupled Ginzburg-Landau equations were used to describe the pulse propagation in the weakly birefringent fibers:

$$\frac{\partial u}{\partial z} = i\beta u - \delta \frac{\partial u}{\partial t} - \frac{ik''}{2}\frac{\partial^2 u}{\partial t^2} + \frac{k'''}{6}\frac{\partial^3 u}{\partial t^3} + i\gamma\left(|u|^2 + \frac{2}{3}|v|^2\right)u + \frac{i\gamma}{3}v^2 u^* + \frac{g}{2}u + \frac{g}{2\Omega_g^2}\frac{\partial^2 u}{\partial t^2},$$

$$\frac{\partial v}{\partial z} = -i\beta v + \delta \frac{\partial v}{\partial t} - \frac{ik''}{2}\frac{\partial^2 v}{\partial t^2} + \frac{k'''}{6}\frac{\partial^3 v}{\partial t^3} + i\gamma\left(|v|^2 + \frac{2}{3}|u|^2\right)v + \frac{i\gamma}{3}u^2 v^* + \frac{g}{2}v + \frac{g}{2\Omega_g^2}\frac{\partial^2 v}{\partial t^2}, \quad (1)$$

where, $u$ and $v$ are the normalized envelopes of the optical pulses along the two orthogonal polarized modes of the optical fiber. $2\beta = 2\pi\Delta n/\lambda$ is the wave-number difference between the two modes. $2\delta = 2\beta\lambda/2\pi c$ is the inverse group velocity difference. $k''$ is the second order dispersion coefficient, $k'''$ is the third order dispersion coefficient and $\gamma$ represents the nonlinearity of the fiber. g is the saturable gain coefficient of the fiber and $\Omega_g$ is the bandwidth of the laser gain. For undoped fibers $g=0$; for erbium doped fiber, we considered its gain saturation as

$$g = G\exp[-\frac{\int(|u|^2 + |v|^2)dt}{P_{sat}}] \quad (2)$$

where $G$ is the small signal gain coefficient and $P_{sat}$ is the normalized saturation energy.

The saturable absorption of graphene materials is described by the following rate equation, which has been previously used for semiconductor saturable absorption mirror[3]:

$$\frac{\partial \alpha_S^*}{\partial t} = -\frac{\alpha_S^* - \alpha_0^*}{\tau} - \frac{|u|^2 + |v|^2}{E_{sat}}\alpha_S^* \quad (3)$$

where $\tau$ is the absorption recovery time, $\alpha_0^*$ is the initial absorption of the absorber, and $E_{sat}$ is the absorber saturation energy. To make the simulation possibly close to the experimental conditions, we used the following fiber and laser parameters: $\gamma=3$ W$^{-1}$km$^{-1}$, $\Omega_g=24$ nm, $P_{sat}=100$ pJ, $k''_{SMF}=$ -23 ps$^2\cdot$km$^{-1}$, $k''_{EDF}=$ -13

ps$^2$·km$^{-1}$, $k'''$= -0.13 ps$^3$·km$^{-1}$, cavity length $L$= 78.6 m. We used the standard split-step Fourier technique to solve the equations and a so-called pulse tracing method to model the effects of laser oscillation, as we discussed before[1].